\documentclass[journal]{IEEEtran}
\usepackage[left=0.75in, right=0.75in, top=1in, bottom=0.9in]{geometry}
\makeatletter
\def\ps@headings{%
\def\@oddhead{\mbox{}\scriptsize\rightmark \hfil \thepage}%
\def\@evenhead{\scriptsize\thepage \hfil \leftmark\mbox{}}%
\def\@oddfoot{}%
\def\@evenfoot{}}
\makeatother 

\usepackage{amsmath,amssymb,mathrsfs,amsthm}
\usepackage{mathtools}
\usepackage{cite}
\usepackage{graphicx,stfloats,subfigure,multicol}%amsmath
\usepackage{epsfig,epsf,psfrag,amssymb,amsfonts,latexsym,graphicx,mathrsfs,subfigure}
\usepackage{bbm}
\usepackage{chngpage}
\usepackage[dvips]{color}
\usepackage{textcomp}
\usepackage{hyperref}
\usepackage{url}
\usepackage[hyphenbreaks]{breakurl}
\usepackage[normalem]{ulem}
\usepackage{algpseudocode}
\newsavebox{\ieeealgbox}

\usepackage{array}
 \def\old#1{}    % Please don't remove this... This command includes the text to be deleted.

\def\nn{\nonumber}
\def\beq{\begin{equation}}
\def\eeq{\end{equation}}
\def\bea{\begin{eqnarray}}
\def\eea{\end{eqnarray}}
\def\ba{\begin{array}}
\def\ea{\end{array}}

\def\bitem{\begin{itemize}}
\def\eitem{\end{itemize}}
\def\ben{\begin{enumerate}}
\def\een{\end{enumerate}}

\def\eg{{\it e.g., \/}}

\def\ie{{\it i.e.,\ \/}}

\definecolor{bgrd}{rgb}{1,1,1}
\definecolor{gray}{rgb}{0.5,0.5,0.5}
\definecolor{dkr}{rgb}{0.7,0.1,0.2}
\definecolor{dkb}{rgb}{0.1,0.1,0.8}

\def\scalefig#1{\epsfxsize #1\textwidth}

\newcommand{\mbbE}{\mathbb{E}}

\def\Dc{{\cal D}}
\def\Ec{{\cal E}}

\def\Hc{{\cal H}}

\def\Kc{{\cal K}}

\def\Qc{{\cal Q}}

\begin{document}

\title{Adaptive Subband Compression for Streaming of Continuous Point-on-Wave and PMU Data}

\author{Xinyi Wang,~\IEEEmembership{Student~Member,~IEEE},
Yilu Liu,~\IEEEmembership{Fellow,~IEEE},
Lang~Tong,~\IEEEmembership{Fellow,~IEEE},
\thanks{\scriptsize  Xinyi Wang and
Lang Tong ({\tt \{xw555,lt35\} @cornell.edu}) are  with the School of Electrical and Computer Engineering, Cornell University,  USA. Yilu Liu ({\tt liu@utk.edu}) is with the Department of Electrical and Engineering and Computer Science, The University of Tennessee, Knoxville.
This work was supported in part by the National Science Foundation under Grants 1932501 and 1816397}}
\maketitle

{
\begin{abstract}
A data compression system capable of providing real-time streaming of high-resolution continuous point-on-wave (CPOW) and phasor measurement unit (PMU) measurements is proposed.  Referred to as adaptive subband compression (ASBC), the proposed technique partitions the signal space into subbands and adaptively compresses subband signals based on each subband's active bandwidth.  The proposed technique conforms to existing industry phasor measurement standards, making it suitable for streaming high-resolution CPOW and PMU data either in continuous or burst on-demand/event-triggered modes.  Experiments on synthetic and real data show that ASBC reduces the CPOW sampling rates by several orders of magnitude for real-time streaming while maintaining the precision required by industry standards.
\end{abstract}
\begin{IEEEkeywords}
Continuous point-on-wave (CPOW) measurement.  Phasor measurement units (PMU).  Subband compression.  Adaptive data compression. Wide-area monitoring systems (WAMS). Real-time monitoring and control.
\end{IEEEkeywords}
}

\section{Introduction} \label{sec:intro}

With deeper penetration of inverter-based resources that exhibit low inertia and fast dynamics, there are growing needs for high-resolution grid measurement and streaming technology  \cite{Liu:20Workshop,Silverstein&Follum:20NAPSI}.    In \cite{Silverstein&Follum:20NAPSI},
Silverstein and Follum make a compelling case that the time-synchronized continuous point-on-wave (CPOW) measurement technology has the potential to address operational challenges in a broad range of grid applications beyond the capabilities of the existing phasor-measurement unit (PMU) technology.  These applications include monitoring geomagnetic disturbances, subsynchronous resonance (SSR), rapid phase-jumps, and high-resolution monitoring of inverter-based distributed energy resources.   The need for high-resolution data to capture new phenomenon such as super harmonics and oscillations has also been articulated in \cite{Liu:20Workshop}.

Anticipating that future wide-area measurement systems (WAMS) will likely include CPOW, PMU, and SCADA devices, this paper proposes a lossy compression technique for the high-fidelity and high-resolution streaming technology in either continuous or on-demand modes.  By high-fidelity, we mean that the signal reconstruction has the accuracy within specifications of  industry standards. By high-resolution, on the other hand, we mean that the source signal is sampled at sufficiently high frequencies to capture higher-order harmonics, wideband interharmonics, and wideband transients.

\begin{figure}[htb]
\begin{psfrags}
\psfrag{x}[c]{\Large $$}
\psfrag{y}[c]{\Large $$}
\psfrag{z}[c]{\Large $$}
\centerline{\scalefig{0.45}\epsfbox{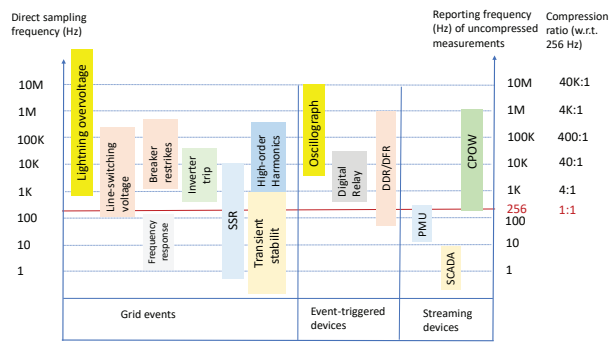}}
\end{psfrags}
\caption{Device sampling frequencies,  (uncompressed) device reporting rates for different applications and devices, and compression ratio required for streaming at the rate of 256 samples/sec. Frequency ranges are approximately illustrated. (Figure adapted from \cite{Silverstein&Follum:20NAPSI,Anderson&Agrawal&vanNess:99book,Perez:10PRE}.)}
\label{fig:sampling}
\end{figure}

CPOW technology produces time-synchronized continuous streams of direct samples of measured signals at a data rate ranging from 256 Hz to 1 MHz \cite{Silverstein&Follum:20NAPSI}.  Point-on-wave (POW) measurement devices already exist, such as digital fault and disturbance recorders (DFR/DDR). Traditionally,   high-resolution point-on-wave (POW) measurements are event-triggered designed for post-event analysis.  Fig.~\ref{fig:sampling} shows the timescales of various grid events and existing measurement devices that generate event-triggered measurements.    Also shown are the two existing streaming technologies based on SCADA\footnote{SCADA: supervisory control and data acquisition.} and PMU devices.

A critical difference between CPOW and PMU/SCADA measurements is that CPOW measurements produce unfiltered high-resolution
voltage and current samples.  Such measurements capture crucial details in transient events that reveal operational risks.  For example, the post-event analysis of the 2016 Big Cut Fire by NERC  \cite{NERC:17} shows that the rapid voltage phase jumps caused many inverter trippings that led to the loss of 1,200 MW solar generation.  Such events would not have been observed with sufficient accuracy by traditional PMU and SCADA measurements.  NERC report concludes that  POW measurements are the most useful sources of data.

Data compression is a key technology for future high-resolution WAMS. If the existing communication infrastructure is used for streaming,
compression ratios of 100-1000 are necessary for monitoring events such as line switching voltages and some of the lightning overvoltages, as shown in Figure~\ref{fig:sampling}. Currently, there is no existing technology that provides such levels of compression.  See  \cite{Tcheou&etal:14TSG}.

Having a high compression ratio is only one measure in evaluating a compression technology.   Fundamental to streaming data compression is the tradeoff among three factors: (i) the compression ratio, (ii) the accuracy of the compression-decompression algorithm,  and (iii) the delay associated with the compression and decompression processes.  The tradeoff between the first two can be formalized in an information-theoretic setting as the {\em rate-distortion tradeoff} by Shannon \cite{Cover&Thomas:Book}.  The last is relevant in streaming applications where encoding/decoding delays are crucial constraints.

\subsection{Related Literature}
There are no compression techniques and standards for the real-time streaming of CPOW data to our best knowledge.  Here we review some of the relevant  technologies that can be candidates for high rate  CPOW/PMU data streaming.

The need for data compression for power system monitoring goes back at least three decades.  Mehta and Russell made one of the earliest contributions in \cite{Mehta&Russell:89TPD}, where they recommended compressing data in the frequency domain using the Fast Fourier Transform (FFT) by discarding high-frequency coefficients. Discarding high-frequency components may lead to significant distortions, however, when the signal has higher-order harmonics.  By processing data
 in blocks, FFT-based techniques introduce inter-block distortions.

%From an information-theoretic perspective, a modification of Mehta and Russell's suggestion by a reverse water-filling strategy can achieve the optimal tradeoff between compression ratio and reconstruction error  \cite{Cover&Thomas:Book}.  The caveat for the optimality of FFT-based compression is that the FFT block size must be large, which conflicts with the low latency requirement of data streaming.  When the FFT block size is limited, the FFT-based compression introduces inter-block distortions, resulting in the loss of signal details and high reconstruction errors.

There is an extensive  literature on compression of POW measurements  by event-triggered digital fault recorders (DFR) \cite{Littler&Morrow:99TPD,Santoso&Powers&Grady:97TPD,Gaouda&etal:00TPD,Santoso&etal:00TPD,Dash&etal:03TPD,Tse&etal:12TPD}.  For such applications, the data sampling rate can be as high as 10MHz.
Because the recorded data are used in post-event analysis, these techniques are designed to be efficient for offline storage instead of real time streaming. To this end, having accurate reconstruction is more important than having a high compression ratio and small compression/decompression delay.  Thus lossless compression techniques are often preferred.  Block linear processing techniques  such as FFT,  discrete-cosine transform (DCT), discrete wavelet transform (DWT) and spline have been developed  \cite{Mehta&Russell:89TPD,Littler&Morrow:99TPD,Santoso&Powers&Grady:97TPD,Gaouda&etal:00TPD,Santoso&etal:00TPD,Dash&etal:03TPD,Tse&etal:12TPD,Meher&Pradhan&Panda:03EPSR,Ning&Etal:11TSG}.   Nonlinear techniques such as neural network, fuzzy logic, and  principal component analysis (PCA) have also been proposed \cite{Meher&Pradhan&Panda:03EPSR,Santoso&etal:00TPD,Ibrahim&Morcos:05TPD,Ge&etal:15TSG}.

PMU data compression for efficient storage has also  attracted considerable attention \cite{Klump&etal:10PESGM,Gadde&etal:16TSG,Kummerow&Nicolai&Bretschneider:18EnergyCon}.   Such applications are significantly different from real-time streaming. It is often assumed that multiple data streams are accessible by the compression algorithm so that spatiotemporal properties can be exploited.  Two-step procedures \cite{Gadde&etal:16TSG,Kummerow&Nicolai&Bretschneider:18EnergyCon} that first compress in the spatial domain using PCA followed by temporal compression techniques (such as DCT and DWT based techniques) have shown to be effective.

The literature on the compression of streaming data for power system monitoring and control is limited.   See a survey on compression techniques for PMU data in real-time smart grid operations  \cite{Tcheou&etal:14TSG}, where the authors reported the capabilities of various compression techniques with the compression ratio up to  5:1 for lossless compressions.   Most relevant to our work are the lossy compressions, categorized by wavelet (and waveform packet) transform techniques, mixed transform, parametric and nonparametric techniques  \cite{Tcheou&etal:14TSG,Das&Sidhu:14TII,Zhang&etal:15TPS}.  These state-of-the-art methods offer 6:1 to 16:1 compression ratios at the normalized mean squared error (NMSE) from -20 to -30 dB.  These techniques typically do not work well for compressing  rapid varying wideband CPOW data.

In a broader context,  the idea of subband compression considered in this paper has long been successfully applied in multimedia communications. Most of the data-streaming techniques (such as the H.264 group) employ some forms of subband compression.  The key to subband compression is to exploit the signal's subband properties to apply high levels of compression in subbands where artifacts of compression are insignificant. For instance, in audio and video compression, the audio/video signals' perceptual properties play a crucial role in achieving tradeoffs among compression ratio, reconstruction accuracy, and encoding-decoding latency.    In this paper, we focus on exploiting the harmonic structure of current/voltage signals for compression.

\subsection{Summary of Results and Contributions}

\begin{figure}[h]
\begin{psfrags}
\psfrag{x}[c]{\Large $$}
\psfrag{y}[c]{\Large $$}
\psfrag{z}[c]{\Large $$}
\scalefig{0.5}\epsfbox{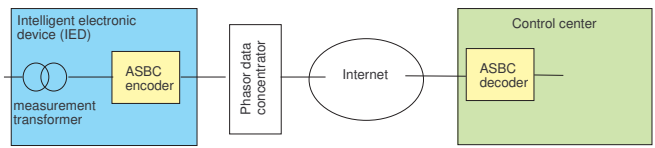}
\end{psfrags}
\caption{An application of ASBC technology for high resolution PMU monitoring of power grids.}
\label{fig:network}
\end{figure}

Given that it is likely that  CPOW and PMU technologies will  coexist in a future wide-area monitoring ecosystem,  it is particularly desirable that a single compression technology applies to both data types.  To this end, the  proposed technique, referred to as {\em adaptive subband compression (ASBC)},  is perhaps the first such compression technique.
Fig.~\ref{fig:network} illustrates a conceptual infrastructure realization of the ASBC technology.
ASBC consists of an encoder for each remote sensing device and a decoder at the fusion center\footnote{A fusion center is a location where data streams from different sensing devices are combined.  A fusion center may be located at  PMU data concentrators (PDC) or the operator's control center.}.  Together, they form the ASBC codec that provides end-to-end data streaming.  Implementations of the ASBC codec are explained in Sec.~\ref{sec:encoder}-\ref{sec:decoder}.

ASBC partitions the signal spectrum into harmonic and interharmonic subbands, as shown in Fig.~\ref{fig:Xf}.  The harmonic subbands are centered at integral multiples of the system frequency (50 or 60 Hz).  Each harmonic subband contains frequency components within the sideband of a specific bandwidth.    The interharmonic subband, on the other hand,  is a set of frequency bands between harmonic subbands \cite{Testa&etal:07TPD}.

A key feature of ASBC is adaptivity.  The encoder monitors activity of subbands and transmits only signals from active subbands.  Such an approach is instrumental when transients and interharmonics are episodic and wideband.   The level of interharmonics may be negligible most of the time and becomes strong suddenly when magnified by resonance. Thus an in-situ compression of interharmonics can achieve a high compression ratio without affecting reconstruction accuracy.

In evaluating the performance of ASBC, we provide a theoretical characterization of the compression ratio and the normalized mean-squared error 
of ASBC  and a set of numerical comparisons between ASBC and selected benchmark techniques.

Key symbols used  are listed in Table~\ref{tab:symbols}.  Otherwise, notations used in this paper are standard.  We use $x(t)$  and $x[n]$  for continuous-time and discrete-time signals, respectively.

{\small
\begin{table}[h]
\caption{\small Major symbols (in alphabetic order).}
\begin{center}
\begin{tabular}{|ll|}
\hline
$e(t)$: &noise outside harmonic subbands.\\
\hline
$F_s$: &sampling frequency of the uncompressed signal\\\hline
$p_k$: & probability of the $k$th subband being active \\
$p_e$: &probability of interharmonic subband being active\\\hline
$R_k$: & rate of quantization for the $k$th subband.\\
$R_e$: &rate of quantization for the interharmonic subband\\\hline
$S_k$:  &downsampling factor of the $k$th subband.\\\hline
$u_k[n]$: & upsampled signal from $\tilde{x}_k^Q[n]$. The data rate is the
\\& same as $y_k[n]$ \\\hline
$x(t),x[n]$: & measurement signal model.\\
$x_k(t),x_k[n]$: & signal component associated with the $k$th harmonic  \\
& with $W_k$ as its bandwidth.\\
$\tilde{x}_k[n]$: &downsampled measurements from $y_k[n]$. \\
$\tilde{x}^Q_k$: &quantized bit-stream of $\tilde{x}_k[n]$ .\\\hline
$\tilde{x}_k[n]$: &downsampled measurements from $y_k[n]$. \\\hline
$y_k[n]$: &  baseband representation of the $k$th harmonic $x_k[n]$.  \\
$\hat{y}_e[n]$: & reconstruction of the interharmonic subband signal. \\\hline
%&FFT/DWT based technique can be used for compression\\
%&of interharmonic subband.\\\hline
$\varepsilon^\chi$: & normalized mean squared error of compression\\
& technique $\chi$\\
$\eta^{\chi}$: &compression ratio of compression  technique $\chi$ \\
\hline
\end{tabular}
%\caption{\small Major symbols (in alphabetic order).}
\end{center}
\label{tab:symbols}
\end{table}
}

%\input intro_v6_k

%\section{Existing standard and technology} \label{sec:standard}
%\input current_v0

\section{Signal model and subband decomposition} \label{sec:model}

We model  the continuous-time voltage (or current) signal $x(t)$  as the sum of $K$ harmonics $x_k(t)$ and an interharmonic component $e(t)$:

\vspace{-1em}
\begin{subequations}\label{eq:model}
\begin{align}
x(t)&=  \sum_{k=1}^K x_k(t) + e(t), \label{eq:xa}\\
x_k(t) &=  a_k(t) \cos(k\Omega_0 t + \phi_k(t)),  \label{eq:xb}
\end{align}
\end{subequations}
where $x_1(t)$ is the signal component associated with the system frequency $F_0$   (\eg $50$ or $60$ Hz),  $\Omega_0=2\pi F_0$,
$x_k(t)$ the $k$th harnomic centered around $kF_0$, and $K-1$  the total number of  higher-order harmonics\footnote{The IEEE Standard C37.118.2-2011 suggests to include higher order harmonics up to the 50th order ($K=50$).}.   Here we allow $x_1(t)$ and its harmonics $x_k(t)$ to  take the general analytical form of (\ref{eq:xb}).  The interharmonic $e(t)$   models  noise outside the harmonic subbands.

\begin{figure}[h]
\begin{psfrags}
\psfrag{w}[c]{$W_0$}
\psfrag{wa}[c]{ $W_1$}
\psfrag{wb}[c]{ $W_2$}
\psfrag{w0}[c]{ $f$}
\psfrag{w1}[c]{ $F_0$}
\psfrag{w2}[c]{ $2F_0$}
\psfrag{w3}[c]{ $3F_0$}
\psfrag{X}[c]{$|X(f)|$}
\scalefig{0.5}\epsfbox{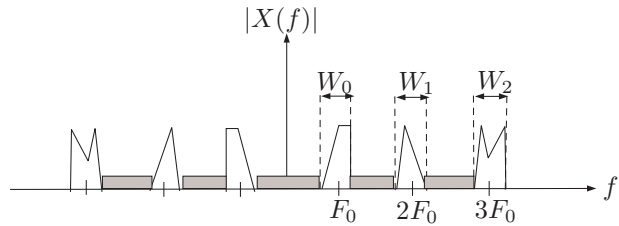}
\end{psfrags}
\caption{\small The spectrum of $x(t)$ and its harmonics.  The grey area represents the spectrum of interharmonics.}
\label{fig:Xf}
\end{figure}

 Let $X(f), X_k(f)$ and $E(f)$  be the Fourier spectra\footnote{Here we assume the existence of Fourier transforms of all signals.}
  of $x(t)$, $x_k(t)$, and $e(t)$, respectively, as illustrated  in Fig.~\ref{fig:Xf}.   We assume that the spectrum $X_k(f)$  of the $k$th harmonic $x_k(t)$  is
centered around $kF_0$  with passband bandwidth\footnote{The {\em passband bandwidth} is defined by the width of the frequency band containing non zero frequency components.} $W_k < F_0$.   The total bandwidth of $x(t)$ is therefore $KF_0+\frac{W_K}{2}\le (K+\frac{1}{2})F_0$.

Sampled at the frequency  $F_s$ (Hz), the discrete-time signal is given by, for $n= 0, \pm 1,  \cdots$,
\begin{subequations}\label{eq:x[n]}
\begin{align}
x[n] &:= x(n/F_s) =   \sum_{k=1}^K x_k[n] + e[n],  \\
x_k[n] &:=  a_k[n] \cos\bigg(k\frac{\Omega_0}{F_s} n + \phi_k[n]\bigg),
\end{align}
\end{subequations}
where $(a_k[n], \phi_k[n])$ are the sampled amplitudes and phase angles, and $e[n]$  is the interharmonic signal.
%Let $X(\omega)=\sum_n x[n] e^{-j\omega n}$ be the discrete-time Fourier transform (DTFT) of $x[n]$.

{Unlike CPOW data, PMU measurements are complex phasors and real frequency measurements that are slowly varying.  To incorporate PMU data model in the same framework, we make a slight generalization of (\ref{eq:model}) by modeling  PMU measurement $x[n]$ as sampled   {\em complex baseband signal} $x(t)$  with $\Omega_0=0$ defined by
\begin{equation}
    x(t)=a_0(t)e^{j\phi_0(t)}.
    \label{eq:Synchrophasor}
\end{equation}
A special form of the above is used by the IEEE C37.118.1-2011 standard for PMU dynamic compliance evaluation\footnote{The IEEE standard C37.118.1-2011 uses the model $x(t)= \frac{X_m}{\sqrt{2}}(1+k_x\cos(\omega t))\angle k_a\cos(\omega t-\pi)$ for dynamic compliance evaluation.}.  With (\ref{eq:Synchrophasor}), the CPOW data compression techniques developed here applies directly to PMU data.}

\begin{center}
\begin{figure*}[t]
\begin{psfrags}
\psfrag{T0}[c]{\small $t=nT_s$}
\psfrag{x}[c]{  $x(t)$}
\psfrag{xn}[c]{  $x[n]$}
\psfrag{e}[c]{  $e^{-j\omega_0n}$}
\psfrag{Ke}[c]{ $\Kc_e$}
\psfrag{K0}[c]{ $\downarrow S_1$}
\psfrag{K1}[c]{ $\downarrow S_2$}
\psfrag{K2}[c]{ $\downarrow S_3$}
\psfrag{K3}[c]{ $\downarrow S_4$}
%\psfrag{Ae}[c]{  $H_e(z)$}
%\psfrag{A0}[c]{ $H_0(z)$}
%\psfrag{A1}[c]{ $H_1(z)$}
%\psfrag{A2}[c]{$H_2(z)$}
%\psfrag{A3}[c]{ $H_3(z)$}
\psfrag{Ae}[c]{  $\Hc_e$}
\psfrag{A0}[c]{ $\Hc_1$}
\psfrag{A1}[c]{ $\Hc_2$}
\psfrag{A2}[c]{$\Hc_3$}
\psfrag{D0}[c]{ $$}
\psfrag{De}[c]{  $\Dc_e$}
\psfrag{D1}[c]{  $\Dc_2$}
\psfrag{D2}[c]{  $\Dc_3$}
\psfrag{D3}[c]{  $\Dc_3$}
\psfrag{Qe}[c]{  $\Qc_e$}
\psfrag{Q0}[c]{  $\Qc_1$}
\psfrag{Q1}[c]{  $\Qc_2$}
\psfrag{Q2}[c]{  $\Qc_3$}
\psfrag{Q3}[c]{  $\Qc_3$}
\psfrag{xe}[c]{    $\tilde{x}_e$}
\psfrag{x1}[c]{    $\tilde{x}_1$}
\psfrag{x2}[c]{    $\tilde{x}_2$}
\psfrag{x3}[c]{     $\tilde{x}_3$}
\psfrag{ye}[c]{    $\tilde{x}^Q_e$}
\psfrag{y1}[c]{    $\tilde{x}^Q_1$}
\psfrag{y2}[c]{     $\tilde{x}^Q_2$}
\psfrag{y3}[c]{     $\tilde{x}^Q_3$}
\psfrag{z}[l]{     $b$ (bit stream)}
\psfrag{ze}[c]{     $y_e$}
\psfrag{z1}[c]{     $y_1$}
\psfrag{z2}[c]{  $y_2$}
\psfrag{z3}[c]{    $y_3$}
\psfrag{z4}[c]{    $y_3[n]$}
\psfrag{ge}[c]{    $(y_e, w_e)$}
\psfrag{g1}[c]{    $(y_2, w_2)$}
\psfrag{g2}[c]{   $(y_3, w_3)$}
\psfrag{g3}[c]{    $(y_3, w_3)$}
\centerline{\scalefig{0.8}\epsfbox{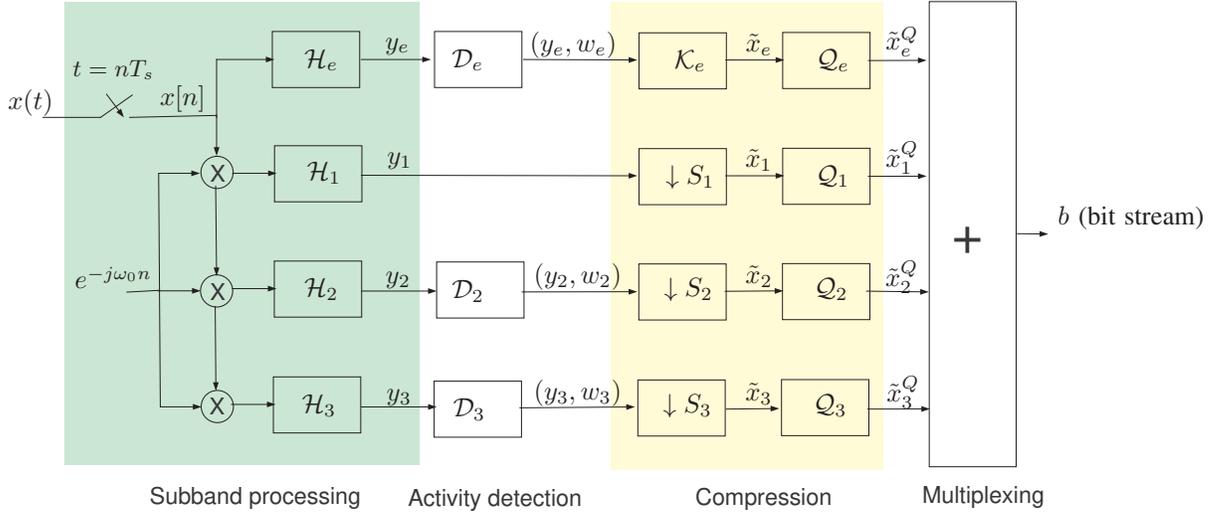}}
\end{psfrags}
\caption{A three-subband ASBC encoder where $h_k[n]$  is the impulse response  of the $k$th subband filter.  Time indices of  signals are obmitted with $y_i$ standing for sequence $(y_i[n])$.}
\label{fig:encode}
\end{figure*}
\end{center}

% \vspace{-2em}
\section{Adaptive Subband Compression: Encoder} \label{sec:encoder}

%
%
%\begin{center}
%\begin{figure*}[b]
%\begin{psfrags}
%\psfrag{T0}[c]{$t=nT_s$}
%\psfrag{x}[c]{ $\hat{x}[n]$}
%\psfrag{R}[c]{ ${\rm Re}$}
%\psfrag{xn}[c]{ $x[n]$}
%\psfrag{e}[l]{$e^{j\omega_0n}$}
%\psfrag{e1}[c]{$e^{j2\omega_0n}$}
%\psfrag{e2}[c]{ $e^{j3\omega_0n}$}
%\psfrag{K0}[c]{$\uparrow S_0$}
%\psfrag{K1}[c]{ $\uparrow S_1$}
%\psfrag{K2}[c]{ $\uparrow S_2$}
%\psfrag{A0}[c]{ $\hat{h}_0[n]$}
%\psfrag{A1}[c]{$\hat{h}_1[n]$}
%\psfrag{A2}[c]{$\hat{h}_2[n]$}
%\psfrag{Ae}[c]{ $\Kc_e^{\dagger}$}
%\psfrag{D0}[c]{ $$}
%\psfrag{D1}[c]{$\Dc_1$}
%\psfrag{D2}[c]{ $\Dc_2$}
%\psfrag{D3}[c]{$\Dc_3$}
%\psfrag{Q0}[c]{$\Qc_0$}
%\psfrag{Q1}[c]{ $\Qc_1$}
%\psfrag{Q2}[c]{ $\Qc_2$}
%\psfrag{Q3}[c]{ $\Qc_3$}
%\psfrag{y1}[c]{ $x^Q_0$}
%\psfrag{y2}[c]{ $x^Q_1$}
%\psfrag{y3}[c]{ $x^Q_2$}
%\psfrag{ye}[c]{ $x^Q_e$}
%\psfrag{z}[c]{ $b$}
%\psfrag{z1}[c]{ $z_0$}
%\psfrag{z2}[c]{ $z_1$}
%\psfrag{z3}[c]{ $z_2$}
%\psfrag{g1}[c]{ $\hat{y}_0[n]$}
%\psfrag{g2}[c]{ $\hat{y}_1[n]$}
%\psfrag{g3}[c]{ $\hat{y}_2[n]$}
%\psfrag{ge}[c]{ $\hat{y}_e[n]$}
%\centerline{\scalefig{0.7}\epsfbox{figs/Schematic_Decode1.eps}}
%\end{psfrags}
%\caption{A three-subband ASBC decoder.  Time indices of internal signals are obmitted with $y_i$ standing for $y[n]$.}
%\label{fig:decode}
%\end{figure*}
%\end{center}

As illustrated in Fig.~\ref{fig:encode}, an ASBC encoder  partitions the signal spectrum into a set of frequency bands,  adaptively masks inactive bands,  and encodes  the unmasked bands in parallel.    These individual components are explained below.

\vspace{-0.5em}
 \subsection{Subband decomposition} From the output of the sensor transformer, the continuous-time measurement signal $x(t)$ is sampled at $F_s$ Hz.  The discrete-time signal $x[n]$ is frequency down-shifted and passed through a  filterbank $\Hc=(\Hc_e,\Hc_1,\cdots, \Hc_K)$ that
 extracts the subband signal $x_k[n]$ in its baseband represention $y_k[n]$.  Specifically, the  output of the $k$th subband filter is a complex time series
\begin{equation}
    y_k[n] = (x[n]e^{-j k\omega_0 n}) \circledast h_k[n],~~\omega_0 := \frac{2\pi F_0}{F_s},
\end{equation}
where $\circledast$ is the convolution operator.  Ideally, the filter for the $k$th subband is a low-pass filter with bandwidth $W_k/2$, whose output $y_k[n]$ is the baseband representation of the $k$th harmonic signal $x_k[n]$.

The interharmonic distortion $y_e[n]$  whose spectrum corresponds to the grey area of power spectrum in Fig.~\ref{fig:Xf} can be  extracted  by
\begin{equation}
    y_e[n] = x[n] - \sqrt{2}{\rm Re}\bigg(\sum_{k=1}^K y_k[n]e^{jk\omega_0n}\bigg).
\end{equation}

In absence of high order and interharmonics, $x(t) = x_1(t)$ in (\ref{eq:model}), and only $y_1[n]$ is non-zero.

\begin{center}
\begin{figure*}[t]
\begin{psfrags}
\psfrag{T0}[c]{$t=nT_s$}
\psfrag{x}[c]{ $\hat{x}[n]$}
\psfrag{R}[c]{ ${\rm \sqrt{2}Re}$}
\psfrag{xn}[c]{ $x[n]$}
\psfrag{e}[l]{$e^{j\omega_0n}$}
\psfrag{e1}[c]{$e^{j2\omega_0n}$}
\psfrag{e2}[c]{ $e^{j3\omega_0n}$}
\psfrag{K0}[c]{$\uparrow S_1$}
\psfrag{K1}[c]{ $\uparrow S_2$}
\psfrag{K2}[c]{ $\uparrow S_3$}
\psfrag{A0}[c]{ $\Hc_1^\dagger$}
\psfrag{A1}[c]{$\Hc_2^\dagger$}
\psfrag{A2}[c]{$\Hc_3^\dagger$}
\psfrag{Ae}[c]{ $\Kc_e^{\dagger}$}
\psfrag{D0}[c]{ $$}
\psfrag{D1}[c]{$\Dc_1$}
\psfrag{D2}[c]{ $\Dc_2$}
\psfrag{D3}[c]{$\Dc_3$}
\psfrag{Q0}[c]{$\Qc_0$}
\psfrag{Q1}[c]{ $\Qc_1$}
\psfrag{Q2}[c]{ $\Qc_2$}
\psfrag{Q3}[c]{ $\Qc_3$}
\psfrag{y1}[c]{ $\tilde{x}^Q_1$}
\psfrag{y2}[c]{ $\tilde{x}^Q_2$}
\psfrag{y3}[c]{ $\tilde{x}^Q_3$}
\psfrag{ye}[c]{ $\tilde{x}^Q_e$}
\psfrag{z}[r]{ $b$ (bit stream)}
\psfrag{z1}[c]{ $u_0$}
\psfrag{z2}[c]{ $u_1$}
\psfrag{z3}[c]{ $u_2$}
\psfrag{g1}[c]{ $\hat{y}_1[n]$}
\psfrag{g2}[c]{ $\hat{y}_2[n]$}
\psfrag{g3}[c]{ $\hat{y}_3[n]$}
\psfrag{ge}[c]{ $\hat{y}_e[n]$}
\centerline{\scalefig{0.7}\epsfbox{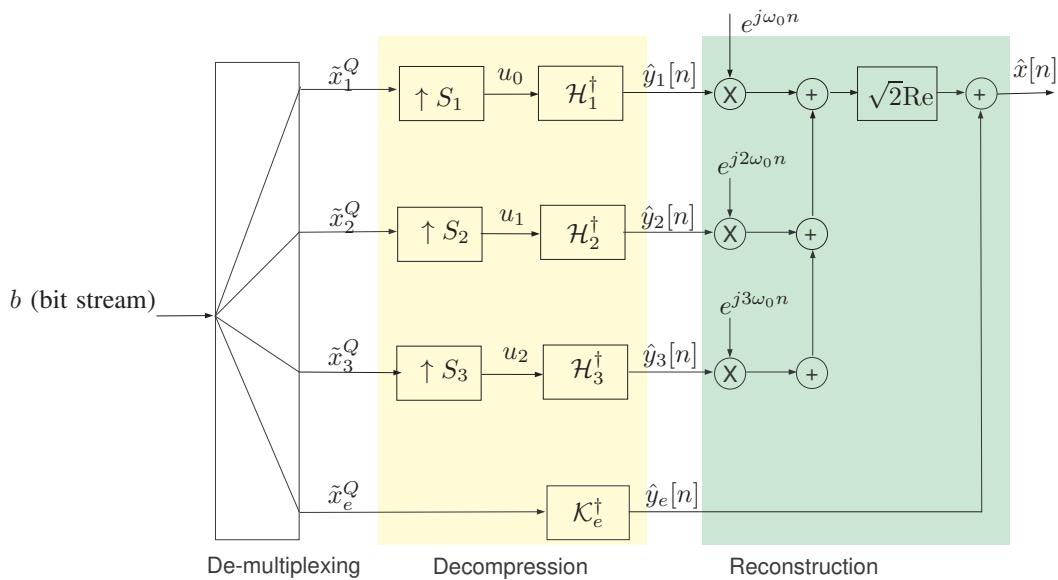}}
\end{psfrags}
\caption{A three-subband ASBC decoder.  Time indices of internal signals $(x_i^Q, u_i)$ are omitted.}
\label{fig:decode}
\end{figure*}
\end{center}

\vspace{-0.5em}
\subsection{Activity detection}  Except $y_1[n]$  from the output of subband filter $\Hc_1$ corresponding to the subband associated with the system frequency $F_0$,  the outputs from the rest of subband filters are passed thorough activity detectors $(\Dc_e, \Dc_k)$  to determine the level of compression required, ranging from  transmitting  at the subband Nyquist rate to full compression that eliminates the transmission of $y_k[n]$.

The activity detection is performed on  blocks of samples. The detector $\Dc_k$  takes a block of samples and outputs an indicator $w_k=1$  for the block  if the subband $k$ is active  and $w_k=0$ otherwise.  The detector for inter-harmonic subband does the same way.   A standard implementation of the activity detector is the energy detector.  More sophisticated techniques such as quickest detection or machine-learning based detection can also be used.

%Detector $\Dc_k$ can be implemented by an energy detector on a block-by-block basis.  For a block of $M_k$ data points starting at $n-M_k+1$  and ending at $n$, let the empirical power level be
%\[
%\hat{p}_k[n] =\left\{\begin{array}{ll}
% \frac{1}{M_k}\sum_{m=0}^{M_k-1} |y_k[n-m]|^2, &~n\equiv 0 (\mbox{mod}~M_k)\\
% \sharp & \mbox{otherwise}
%\end{array}\right.,\]
%where $\sharp$ stands for ``no-value'' that is to be ignored downstream.  Detector $\Dc_k$ compares $\hat{p}_k[n]$ with a threshold $\tau$ and produces a masking function for the data block
%\bea
%z_k[n] &=& \left\{\begin{array}{ll}
%1 & \mbox{if $\hat{p}_k[n] \ge  \tau_k $, $n\equiv 0~(\mbox{mod}~M_k)$}\\
%0 & \mbox{if $\hat{p}_k[n] <  \tau_k $ and $n\equiv 0~(\mbox{mod}~M_k)$}\\
%\end{array}\right.\nn\\
%z_k[n-i] &=& z_k[n]~\mbox{for $n\equiv 0 (\mbox{mod}~M_k), 0\le i <M_k$}.\nn
%\eea
%The threshold $\tau_k$ is chosen to control the false positive rate when there is substantial energy in the $k$th harmonic while the detector $\Dc_k$ declares otherwise.

\vspace{-0.5em}
\subsection{Subband compression} The compression of the harmonic subband $k$ is achieved by down-sampling of $y_k[n]$ by $S_k$ fold.  By the (passband) Nyquist sampling theorem, if the $k$th subband has passband bandwidth of $W_k$, then the signal in the $k$th subband can be perfectly reconstructed by sampling $x_k(t)$ at frequency of $W_k$ (Hz).  Given that $x(t)$ is sampled at $F_s$ (Hz), the rate of down-sampling  $S_k$ is given by
\begin{equation}
    S_k =\Bigg\lceil \frac{F_s}{W_k} \Bigg\rceil.
\end{equation}
 If  subband $k$ is active, the down-sampler gives the compressed data sequence
\beq
\tilde{x}_k[n]=\left\{\begin{array}{ll}
y_k[nS_k], & ~\mbox{$w_k[n]=1$,}\\
\sharp, & \mbox{otherwise,}\\
\end{array}\right.
\eeq
where  $\sharp$ is a masking symbol indicating that the data  sample needs not be encoded and transmitted.  The data rate associated with $\tilde{x}_k[n]$ is at most $1/S_k$ of that of $y_k[n]$.

The interharmonic band is expected to be active infrequently. When an interharmonic signal needs to  be transmitted to the control center, an FFT-based (\eg FFT-$(k,L)$  described in Sec.~\ref{sec:FFT}) or wavelet based compression scheme can be used.
See references in \cite{Tcheou&etal:14TSG}.

\vspace{-0.5em}
\subsection{Quantization and multiplexing}   The down-sampled data streams are quantized by  quantizer $(\Qc_e,\Qc_k)$  that maps subband stream $\tilde{x}_k[n]$ into a bit-stream  $\tilde{x}^Q_k$ of $R_k$ bits/sample. A scaler quantizer such as pulse-code modulation (PCM)
quantizes individual sample of $\tilde{x}_k$ into $R_k$ bits of $\tilde{x}^Q_k$, whereas a vector quantizer such as code excited linear prediction (CLEP) or K-mean clustering takes a block of $M$ samples of and quantizes them into a block of  $MR_k$ bits of $\tilde{x}_k^Q$.   The bit-streams from subbands are multiplexed into a single bit-stream $b$ to be delivered to the receiver.
Also communicated the length of inactivity masks  for each block that is not transmitted due to inactivity.

\section{Adaptive Subband Compression: Decoder} \label{sec:decoder}

ASBC decoders are located at regional data concentrators or the control center where compressed streaming data are reconstructed.
Fig.~\ref{fig:decode} illustrates the schematic of an ASBC decoder.  The functionalities of individual components are explained below.

The de-multiplexing block is the inverse of the multiplexing block at the encoder.  It parses the single bit-stream into subband data streams $\tilde{x}^Q_k$ and $\tilde{x}_e^Q$  sent by the transmitter.

The decompression block reverses the compression block and
generates estimated harmonics  (in baseband) $\hat{y}_k$ in two steps.  First,  $\tilde{x}^Q_k[n]$ is up-sampled (interpolated) with zeros (including replacing masked symbols with zeros) to generate sequence $u_k[n]$ that has the same data rate as that of $y_k[n]$.  The interpolated sequence $u_k[n]$ is passed through a subband interpolation  filter $\Hc_k^{\dagger}$ with impulse response $\hat{h}_k[n]$ to produce an estimate of the baseband representation of the $k$th harmonic signal $\hat{y}_k[n]$.  The subband  interpolation filter may be chosen as the matched-filter $\hat{h}_k[n]=h_k[-n]$ to maximize the signal-to-noise ratio.  Other implementations, such as windowed low-pass filters, can also be used.  The decompression of the interharmonic signal directly follows the compression algorithm used at the encoder.

The final decompression step takes the subband signals $\hat{y}_k[n]$  produce an estimate of  the original direct sampled $x[n]$ of $x(t)$ in the encoder:
\begin{equation}
    \hat{x}[n]=\sqrt{2} {\rm Re} \Bigg(\sum_{k=1}^K \hat{y}_k[n]e^{j k \omega_0 n} \Bigg) + \hat{y}_e[n].
\end{equation}

\section{Rate-Distortion Characteristics} \label{sec:performance}

The standard measure of lossy compression is the rate-distortion curve that highlights the tradeoff between the level of compression and the accuracy of the reconstruction.  A well designed compression scheme has a monotonic rate-distortion curve: the higher the rate of the compressed signal, the lower the compression ratio,  the lower the distortion.

 In this paper, we adopt the compression ratio and  the normalized mean-squared error to characterize the rate-distortion characteristics.
Given a compression scheme $\chi$,  its {\em compression ratio\footnote{The theoretical compression ratio is defined by excluding protocol overhead that  has sublinear growth with the data length.}} defined by
\beq \label{eq:CR}
\eta^{\chi} =  \frac{R^{\text{uc}}}{R^{\chi}},
 \eeq
where $R^{\text{uc}}$  is the data rate (bits/sec) of the uncompressed signal and  $R^{\chi}$ the rate of the compressed stream.

Let $x[n]$ be the original (uncompressed) signal and $\hat{x}^{\chi}[n]$ the reconstructed signal at the decoder.
The {\em normalized mean-squared error} (NMSE) in (dB) is defined  by
\bea
    \Ec^{\chi} &=&10\log_{10} \frac{\sum_{n=1}^{N}|x[n]-\hat{x}^{\chi}[n]|^2}{\sum_{n=1}^{N}x^2[n]}~~(\mbox{dB})\nn\\
    &\rightarrow& 10\log_{10} \frac{\mbbE(|x[n]-\hat{x}^{\chi}[n]|^2)}{\mbbE(|x[n]|^2)} =(\text{\rm SNR}^{\chi})^{-1},    \label{eq:NMSE}
\eea
where $N$ is the length of the data sequence, and the mean-square convergence of (\ref{eq:NMSE}) assumes regularity conditions.
Note that  $\frac{1}{\Ec^{\chi}}$  has the interpretation to be the  signal-to-reconstruction noise ratio (SNR).

For the application at hand, the data rate of the uncompressed data stream can be measured by
\begin{equation}
    R^{\text{uc}}=F_sR_Q~ \mbox{(bits/sec)},
\end{equation}
where $F_s$ represents the sampling frequency of the measured signal $x(t)$ and $R_Q$  the rate of quantization (bits/sample).
The distortion of the uncompressed scheme comes only from quantization error.  With  $R_Q$ bits  PCM quantization,  the NMSE is approximately  by
\begin{equation}
    \Ec^{\text{uc}} \approx  -6R_Q  + 1.25~~({\rm dB}).
\end{equation}

\subsection{Rate-distortion measure of  ASBC: $(\eta^{\text{ASBC}},\Ec^{\text{ASBC}})$}
We provide a characterization of the compression ratio $\eta^{\text{ASBC}}$ and the NMSE of the reconstruction $\Ec^{\text{ASBC}}$.

The data rate of the  compressed data stream by ASBC is
\begin{equation}
    R^{\text{ASBC}}= \sum_{k=1}^{K} p_k\frac{F_s}{S_k}R_k + p_e F_s R_e~ \mbox{(bits/sec)},
\end{equation}
where $F_s$   is the sampling frequency of the uncompressed data, $p_k$ the probability of $k$th subband being active, $R_k$  the rate (bits/sample) of the quantization  in  the $k$th subband, $S_k$ the down-sampling rate of $k$th subband, $p_e$ the probability that the interharmonic subband is active, and $R_e$ the rate of quantization of the  interharmonic subband.     The compression ratio of ASBC is given by
\begin{equation}
        \eta^{\text{ASBC}} = \frac{R^{\text{uc}}}{R^{\text{ASBC}}}=\left(\sum_{k=1}^{K}p_k\frac{R_k}{R_QS_k} + p_e \frac{R_e}{R_Q}\right)^{-1}.
    \label{CompressionRate}
\end{equation}
If we ignore quantization, ASBC gains via adaptively down-sampling of subband signals based on subband activity.  As an example, for the 6kHz sampling of the original signal and a harmonic subband of 6Hz bandwidth, ASBC achieves  the compression ratio $1000:1$ for that subband.

The NMSE measure $\Ec^{\text{ASBC}}$ of ASBC depends on how accurately ASBC can detect the activities of harmonic and interharmonic subbands. Assuming all harmonic subbands are active and there is no interharmonics, \ie $p_e=0$, we have $\Ec^{\text{ASBC}} \approx \Ec^{\text{uc}}$ because ASBC achieves perfect reconstruction of each harmonic signals by the Nyquist sampling theorem.  In practice, $\Ec^{\text{ASBC}} > \Ec^{\text{uc}}$ when  false negative detection occurs or when there is interharmonic signal.

\subsection{Rate-distortion measure of FFT-$(k,L)$ : $(\eta^{\text{FFT}},\Ec^{\text{FFT}})$} \label{sec:FFT}
A benchmark compression scheme is based on the fast Fourier transform (FFT), herein referred to  as FFT-$(k,L)$.  It  takes a block of $L$ data samples, computes the FFT coefficients, and keeps only the $k$ largest coefficients {$X_{j}^{(1)},\cdots,X_{j}^{(k)} $} (corresponding to the positive frequencies in the $j^{th}$ block) and masks the rest.

The compression ratio of FFT-$(k,L)$  is given by
\begin{equation}
    \eta^{\text{FFT-$(k,L)$}} = \frac{L}{k},
\end{equation}
where we ignore the $\log_2 L$ bits needed to encode the frequency locations.
The NMSE of FFT-$(k,L)$  is given by
\begin{equation}
    \Ec^{\text{FFT-$(k,L)$}} = 10\log\bigg( 1-\frac{\sum_{j=1}^{N/L}\sum_{i=1}^{k} |{X_{j}^{(i)}}|^2}{\sum_{i=1}^N |x[i]|^2}\bigg).
\end{equation}

\section{Numerical Results}\label{sec:simulation}

We present numerical results in two categories using synthetic and real data. The first category is the compression of CPOW measurements, where we studied the compression of actual CPOW measurements and synthetic waveforms with characteristics of power system signals.  The tests using synthetic waveforms allowed us to evaluate the performance of benchmark techniques under different scenarios of transient events.   We also studied the compression of  directly  sampled voltage measurements at the sampling frequency of 6 kHz.  This was a case that the signal has significant harmonics and interharmonics.  The second category is the compression of PMU measurements, including synchrophasor measurements and frequency estimates, where both synthetic and actual PMU data were used for experiments.

ASBC was compared with four benchmark techniques:  (i) FFT-based compression described in Sec.~\ref{sec:FFT},  (ii) Multi-resolution discrete wavelet transform (DWT) using the  same parameter (wavelet function, decomposition level) selection method as in \cite{Ning&Etal:11TSG}, (iii) Compressive sampling (CS) \cite{Das&Sidhu:14TII}, and (iv) Exception compression with swing door trending compression (EC-SDT) \cite{Zhang&etal:15TPS}.    Rate-distortion curves that  plot reconstruction error against compression ratio $\eta$ are used in the compression.   We use  NMSE (dB) to measure {\em reconstruction error} for CPOW data compression  and the maximum Total Vector Error (TVE)  defined in \cite{IEEEStdC37:11} for PMU compression.

\subsection{Compressions of synthetic CPOW measurements} \label{sec:synthetic}
The purpose of this experiment was to test the performance of compression when the regular sinusoidal waveform was interrupted by episodes of transient events with signals of abnormal characteristics.   We focused on four types of event signals shown in Table~\ref{table:waveform}.
Among the set of waveforms at the event state, the amplitude modulation (AM) and frequency modulation (FM) waveforms were designed according to the requirement on performance under dynamic compliance specified by \cite{IEEEStdC37:11}.  The linear chirp waveform was used to simulate the frequency ramping events, and the interharmonic (IH) waveform modeled frequency components not in multiples of the system frequency $F_0$.

\begin{table}[h]
\renewcommand{\arraystretch}{1.3}
  \caption{Test waveforms}
    \vspace{0.5em}
    \begin{tabular}{c||c}
        \hline
         \bfseries Test cases & \bfseries Signal waveforms \\
         \hline\hline
        Normal state: & $x_N(t)= \sum_{k=0}^K  a_k \cos(2\pi60(k+1)t+\phi_k)$   \\
        ~~~$\sigma_t=0$  & \\  \hline
        Event:   &   $x^{\text{AM}}(t)=\alpha_0 +\alpha \cos(2\pi \Delta t)\cos(2\pi60t + \theta_0)$ \\
        ~~~$\sigma_t=1$           &   $x^{\text{FM}}(t)=\alpha \cos(2\pi 60t+\beta \cos(2\pi \Delta t-\pi)+\theta_0)$    \\
        &  $x^{\text{Chirp}}(t) = \alpha \cos(2\pi(59.5+\gamma t) + \theta_0)$   \\
                & $x^{\text{IH}}(t) =  \sum_{k=1}^W \alpha_k\cos(2\pi f_k t+\theta_k)$
        \\\hline
    \end{tabular}

    \vspace{0.5em}
    { Signal parameters: $\sigma_t$--the switching state, $(a_k,\phi_k)$--amplitude and phase angles of harmonics, $(\alpha_k,\beta,\gamma,\theta_k)$--transient parameters, and $f_k$--interharmonic frequencies.}
    \label{table:waveform}
\end{table}

To simulate transient events, we used a two-state Markov switching model that modulated the signal between the normal state ($\sigma_t=0$)  waveform $x^{\text{Norm}}(t)$ and the event state ($\sigma=1$) with waveforms chosen from AM, FM, chirp, and IH signals. The Markov switching process was characterized by state transition rate  $(\lambda,\mu)$ where $1/\lambda$ was the expected holding time of the normal state and $1/\mu$  the expected holding time of the event state.

\begin{figure}[h]
    \centering
    \includegraphics[width=1.6in]{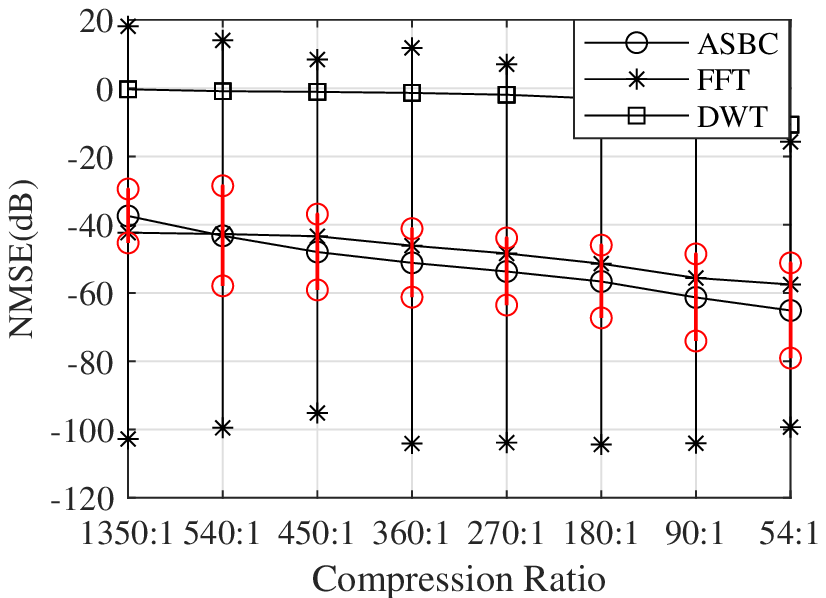}
    \includegraphics[width=1.6in]{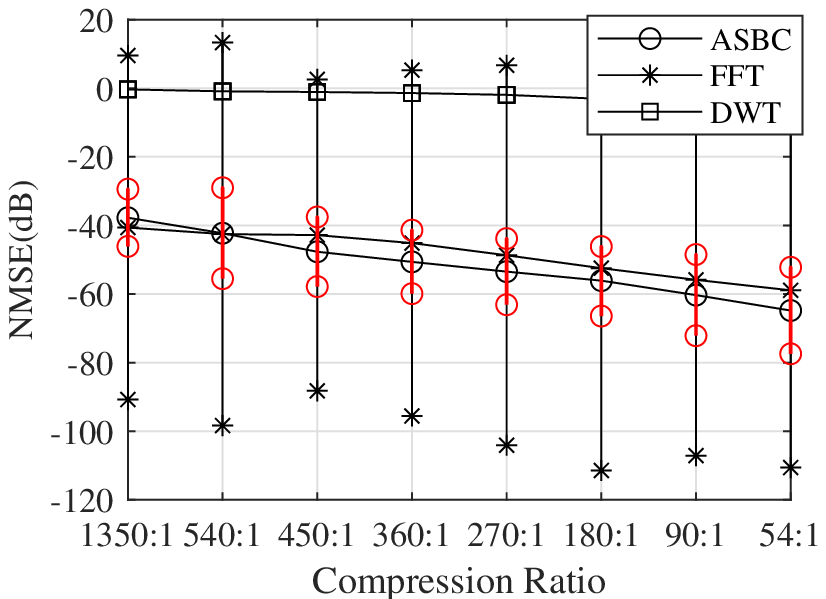}\\
         \includegraphics[width=3in]{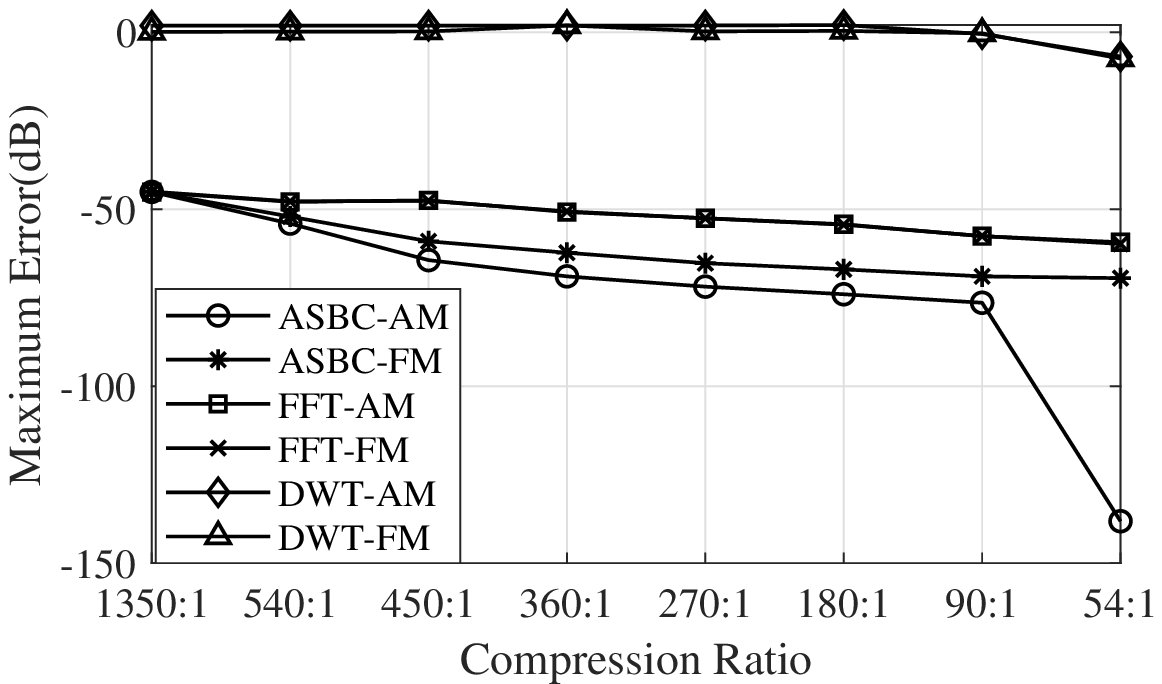}
    \caption{{\small FM/AM parameters: $\alpha_0=1,\lambda=0.1,\mu=2$.  Top panel  NMSE with 95\% confidence interval for the AM (Left) and FM (right) events.
    No high order harmonics and interharmonics.
Bottom: the maximum reconstruction error for the FM and AM events.}}
    \label{fig:CPAM-0.5}
\end{figure}

Fig.~\ref{fig:CPAM-0.5} shows the rate-distortion curve of ASBC, FFT, and DWT techniques for the FM and AM events.  The sampling rate of the original signal was 5400 Hz.  As shown in the upper panel,   the NMSE of DWT scheme was about 30 to 40 dB higher than those of  FFT and ASBC.   Note that  DWT uses non-sinusoidal basis functions.  They tend to be effective for approximately constant or staircase valued signals (such as phasor measurements).  For the compression of sinusoidal voltage/current CPOW measurements,  sinusoidal basis functions used in ASBC and FFT had the advantage of being closer to the native voltage/current waveforms even for linear chirp signals.  In this experiment,  the requirement of high compression ratios forced DWT to discard too many parameters, resulting in high reconstruction error.

\begin{figure}[h]
    \centering
    \includegraphics[width=1.5in]{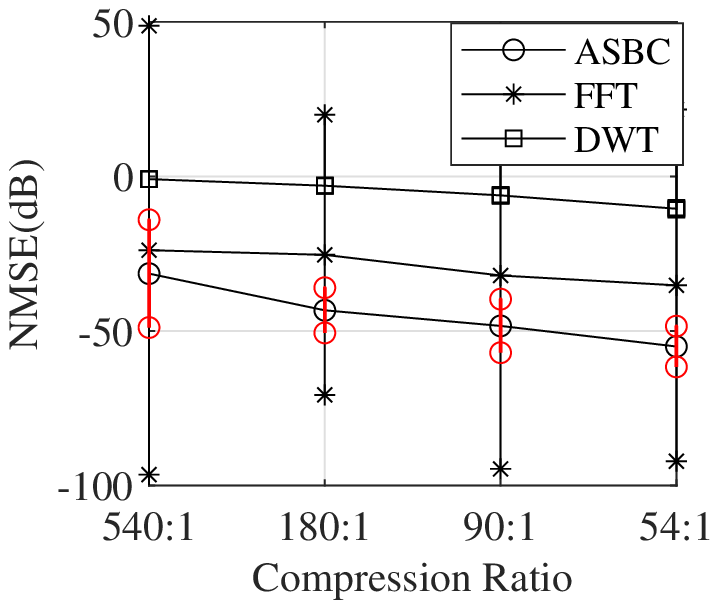}
        \includegraphics[width=1.5in]{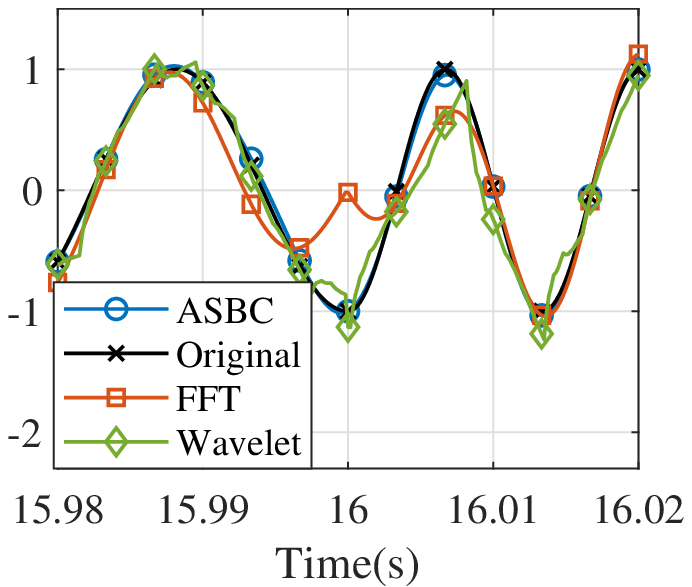}
    \caption{{Left: comparison of rate-distortion curve for linear chip events. Right: a segment of the original and its reconstruction.  $\lambda=0.1$, $\mu=1$}}
    \label{fig:CPLC-0.5}
\end{figure}

ASBC had about 4-6 dB gain over FFT for the compression ratio between 540:1 and 54:1, whereas FFT had 3 dB gain over ASBC at a compression ratio of 1350:1.  Note that the 95\% confidence interval of ASBC was considerably smaller than that of FFT, indicating that the errors of the FFT scheme were more dispersed.  The same behavior was confirmed by the maximum reconstruction error plot at the lower panel of Fig.~\ref{fig:CPAM-0.5}.    The main reason that ASBC outperformed FFT was that FFT introduced discontinuities at the boundaries of FFT blocks. In contrast, the ASBC encoder did not have discontinuities.  This phenomenon was more pronounced for the linear-chirp test cases shown in Fig.~\ref{fig:CPLC-0.5}

Fig.~\ref{fig:CPLC-0.5} shows the rate-distortion plot for the linear chirp events.  The linear chirp signals have a much wider bandwidth, and the achievable compression ratio significantly lower than possible in  FM/AM events.  The left panel shows the NMSE for the three techniques.  Again, DWT was not competitive against  FFT and ASBC techniques,  and ASBC had a considerable gain over the FFT  compression in the low compression ratio regime.  In particular, ASBC had about 15 dB lower NMSE at the compression ratio of 180:1 and 20 dB lower at the compression ratio of 54:1.   The right panel shows the original signal and reconstructed ones in the time domain.  Notice that the normal sinusoidal waveform transitioned to a linear chip at time $t=16$ seconds.  The reconstruction of FFT and DWT around $t=16$  showed a significantly larger error than that of ASBC.

\subsection{Compressions of CPOW voltage measurements.}
\begin{figure}[h]
    \centering
    \includegraphics[width=3in]{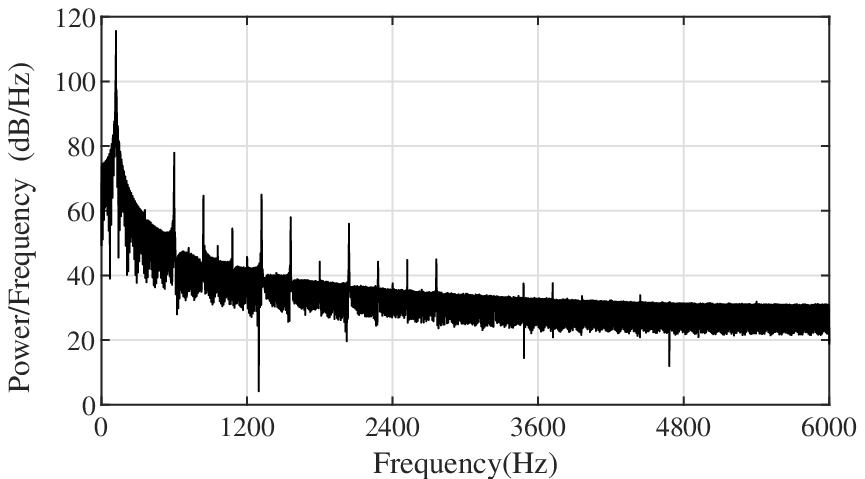}\\
    \includegraphics[width=1.6in]{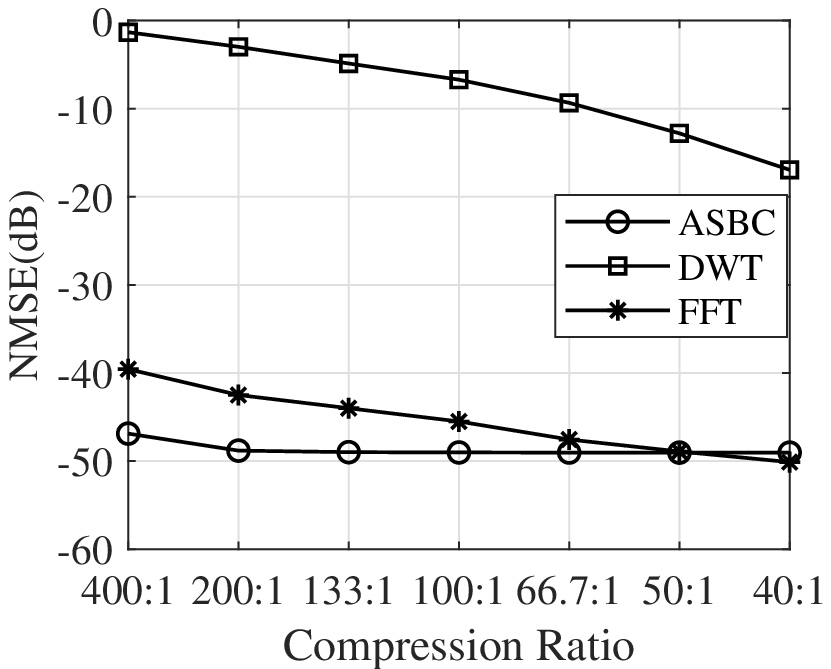}
        \includegraphics[width=1.6in,height=1.3in]{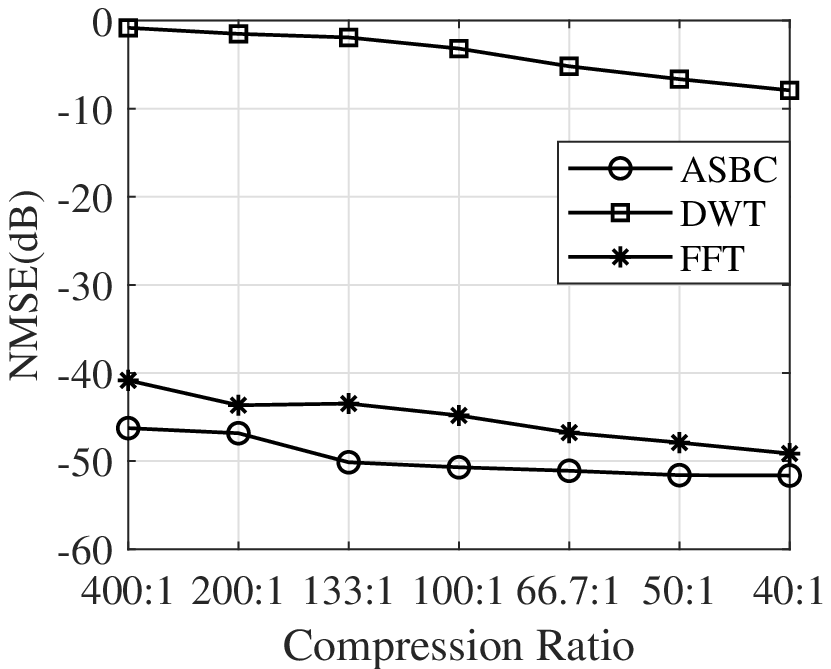}
    \caption{{Top: Power spectrum density of  the direct voltage measurements. Bottom left: comparison of rate-distortion curves without interharmonics. Bottom right: comparison with added interharmonics.}}
    \label{fig:Spectrum}
\end{figure}

We applied ASBC directly to a data set (hereafter referred to as UTK6K) provided by the University of Tennessee, Knoxville. The UTK6K data set consisted of 1.8 million voltage measurements sampled at 6KHz. Fig.~\ref{fig:Spectrum} (Top) shows the power spectrum of the directly sampled data stream, from which we observed the presence of harmonics and interharmonics.  The plot also showed that the energy levels from the 20th to 50th subbands were negligible.

%\begin{figure}[h]
%    \centering
%    \includegraphics[width=3.3in]{figs/UTK6k.eps}
%    \caption{Voltage Compression of 3Hz Subband Width}
%    \label{fig:UTK6KHz3}
%\end{figure}

ASBC was implemented with 3 Hz bandwidth for subbands associated with all 50 harmonics. Only the top  $k$ subbands with the highest energy level were compressed and delivered where $k$ was chosen to have the required compression ratio.  The bottom left panel of Fig.~\ref{fig:Spectrum} shows the rate-distortion curve of ASBC, FFT, and DWT for the compression ratio from 400:1 to 40:1.   For this range of compression ratios, DWT was not competitive.  ASBC was seen to out-perform FFT in the compression ratio range of 400:1 to 100:1, and the two schemes are comparable for the range of 66.7:1 to 40:1.   The reason that FFT-based compression did not perform well was, again,  that the block implementation of FFT introduced discontinuities, which caused reconstruction errors.  As the compression ratio decreased, more FFT coefficients were preserved, the reconstruction error of FFT improved.

To evaluate the effects of interharmonics, we added additional interharmonic transient events to the original UTK6KHz dataset in the same way as experiments discussed in Sec.~\ref{sec:synthetic}.  The bottom right panel of Fig.~\ref{fig:Spectrum} shows the rate-distortion plot with interharmonics subband activated. An energy detector was used to determine when and whether the interharmonics subband should be activated. Only those harmonics subbands with sufficient energy levels were compressed and transmitted. Interharmonics subband, when detected being active by the energy detector, was compressed dynamically to the effective bandwidth ranging from $60$ to $120$Hz.  As shown in the bottom left panel, the presence of interharmonics increased NMSE slightly for ASBC at the high compression ratio.  Overall, ASBC consistently performed better than other methods. The standard deviations were small for all three methods; thus, the confidence intervals were not shown in the plot.

%\begin{figure}[h]
%    \centering
%    \includegraphics[width=2.5in]{figs/InterHar.eps}
%    \caption{Rate Distortion Plot w/ Interharmonics Subband Activated}
%    \label{Fig:InterHar}
%\end{figure}

\subsection{Compressions of PMU dynamic compliance performance}
ASBC can also be used for compressing the streaming of PMU data.  Besides  ASBC,  FFT,  and DWT compressions, we also considered two methods based on compression principles other than Fourier or wavelet-based techniques.  One technique was the application of the compressive sampling (CS) technique \cite{Das&Sidhu:14TII}; the other was based on a combination of exception compression and swing-door-trending compression (EC-SDT).  EC-SDT is a heuristic that forms a time-varying band that covers measurements and interpolating these measurements by a piecewise linear reconstruction.

\begin{figure}[h]
    \centering
    \includegraphics[width=1.65in]{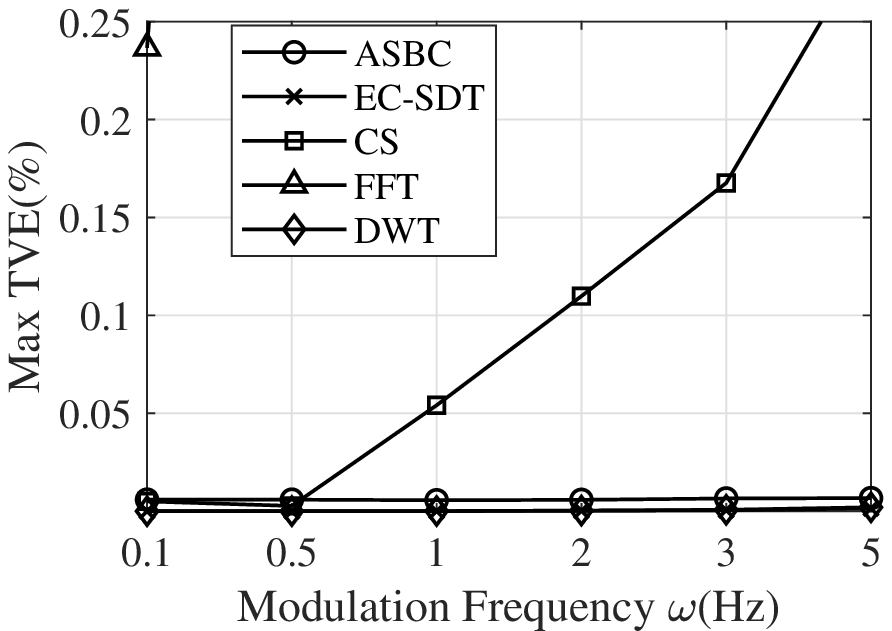}
            \includegraphics[width=1.65in]{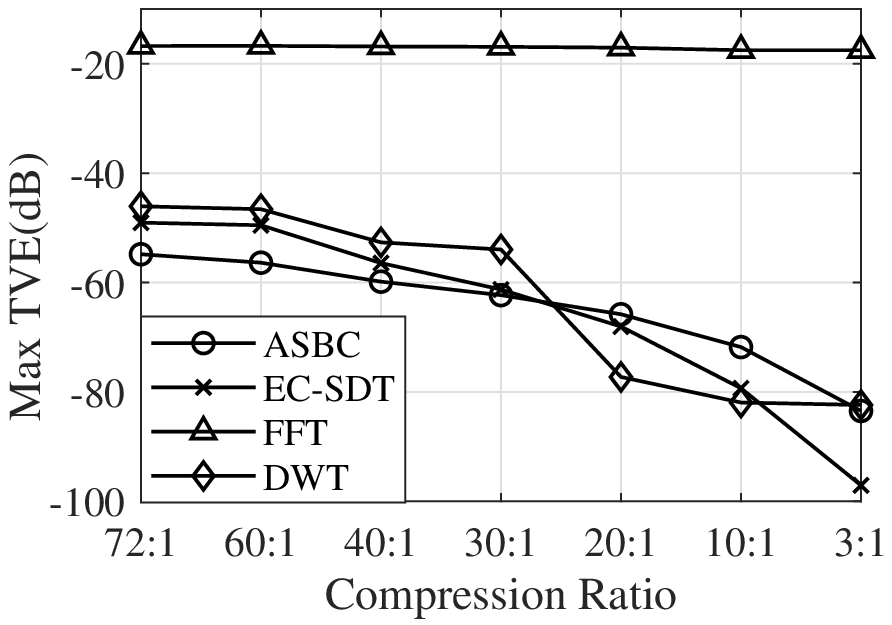}
    \caption{Left: Max TVE fixed at compression ratio 3:1. Right: Rate distortion with $\omega=5$Hz. $k_x=0.1$, $k_a=0.1$}
    \label{fig:Synthetic-PMU}
\end{figure}

Fig.~\ref{fig:Synthetic-PMU} shows the performance of the dynamic compliance test of the benchmark techniques.  The left panel of Fig.~\ref{fig:Synthetic-PMU} shows the maximum total variation error (TVE) vs. the frequency of amplitude/phase variation at the low compression ratio\footnote{The maximum compression ratio  used in \cite{Zhang&etal:15TPS} was converted to the standard definition in (\ref{eq:CR})} of 3:1.  All tested methods except FFT satisfied the 1\% maximum TVE requirement of  IEEE C37.118.1-2011 up to 2Hz modulation frequency. FFT performed the worst for it suffered badly from the discontinuity between blocks.  The compressive sampling (CS) solution appeared to be more sensitive to the modulation frequency. Among the rest techniques, DWT performs the best at this compression ratio, and EC-SDT performed slightly better than ASBC at the lower modulation frequency.

The right panel of Fig.~\ref{fig:Synthetic-PMU} shows the performance comparison in a  significantly higher range of compression ratios.  ASBC, DWT and EC-SDT performed similarly up to 30:1 compression ratio, and ASBC outperformed both DWT and EC-SDT above the compression ratio of 30:1. The max TVE of FFT remains high throughout the tests due to discontinuity.

\subsection{Compressions ov  PMU frequency measurements}
\begin{figure}[h]
    \centering
   \includegraphics[width=1.6in]{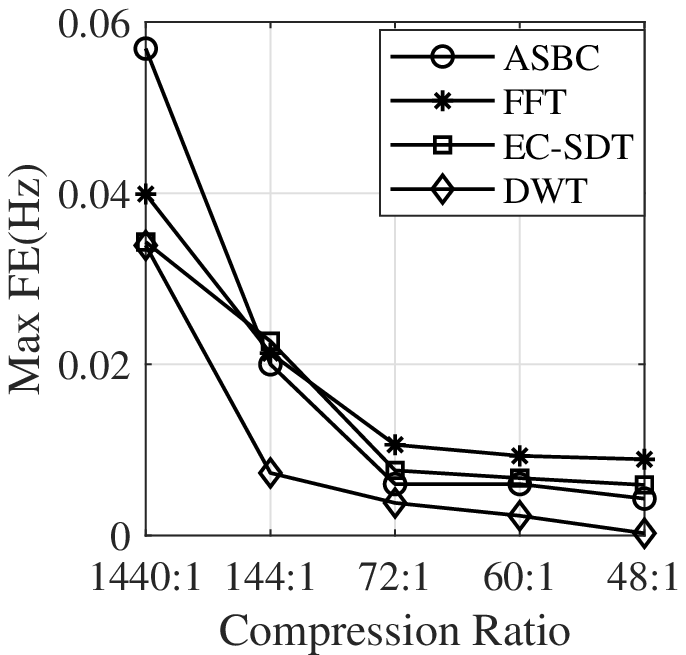}
          \includegraphics[width=1.6in]{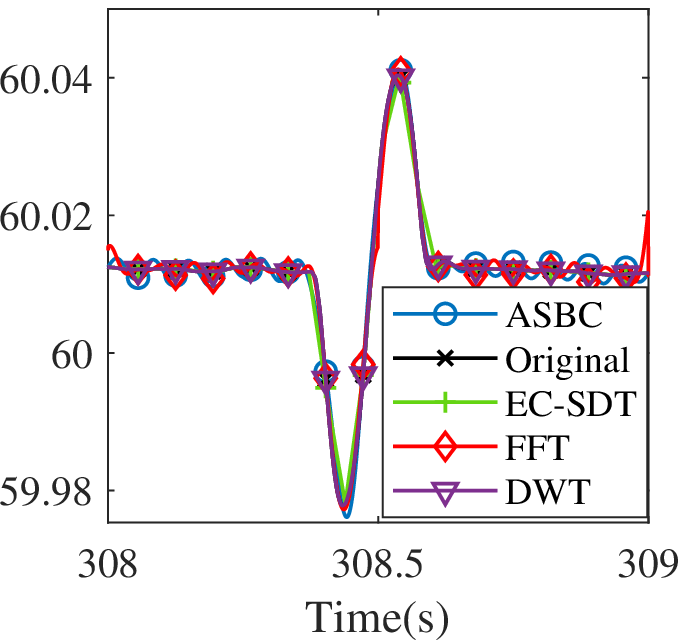}
    \caption{{Maximum FE and time-domain reconstruction}}
    \label{fig:FE}
\end{figure}

We applied ASBC to a dataset referred to as UTK1.44, which consisted of frequency estimates from the  University of Tennessee, Knoxville.  The dataset contained two data streams, each with 1,800,000 samples at the rate of 1440 samples/sec.   One distinct feature of this dataset was the frequency ramping event between 308.4 and 308.6  seconds, as shown in the right panel of  Fig.~\ref{fig:FE}.
As a time series, the frequency measurements are close to being constant at around 60 (Hz).  Thus only a single subband is needed for ASBC. We varied the subband bandwidth to achieve different compression ratios.

{
The performance of the four data compression methods was evaluated base on the maximum error of the reconstruction of the frequency measurements (Max FE) as  defined by the IEEE Standard C37.11  \cite{IEEEStdC37:11} that specifies the acceptable performance is to have Max-FE below 0.005 Hz.  At the compression ratio of 48:1, only ASBC and DWT met the  0.005 Hz threshold.  DWT performed the best for compression ratios above 48:1 because the frequency estimates were approximately constant outside the event around 308.5 sec.
}

{The right panel of Fig.~\ref{fig:FE} showed a segment time-domain reconstructions at the compression ratio of 48:1.    As Fourier based compression techniques, ASBC and FFT compression exhibited small oscillatory error patterns even when the original frequency estimates were approximately constant. Such artifacts are results of the optimal sinusoidal approximation of non-sinusoidal waveforms\footnote{As suggested by the Chebyshev alternation theorem, the optimal approximation error must oscillate around the true value.}.  Because the magnitude of the errors are well within the specification of the IEEE  C37.11 Standard, such artifacts should be treated as noise.
The small spike of reconstruction error of FFT around $t=309$ sec was caused by the discontinuity of the block implementation. }

{
\subsection{Implementation Issues}
We now discuss briefly implementation issues of the proposed compression technology.   As shown in Fig.~\ref{fig:encode}-\ref{fig:decode}, major costs of  implementations are the frequency up/down shifts and subband  filters in the ASBC encoder and decoder. Such operations are standard in communication systems; only minor modifications of the off-the-shelf technology are necessary.   The costs of hardware implementations are low.}

{The overall performance of the compression technology depends, naturally, on setting design parameters appropriately for power system signals.  This includes choosing appropriately the size of the encoding and decoding filterbank to tradeoff implementation accuracy and filtering delays.  Classical signal processing techniques such as windowing offer practical ways to achieve good tradeoffs.
} 

\section{Conclusion}\label{sec:conclusion}
As an emerging technology, real-time and high-resolution CPOW and PMU monitoring of the power grid has the potential to provide situational awareness crucial to reliability and resiliency in the advent of large-scale integration of highly dynamic inverter-based energy resources such as wind, solar, and storage.  As pointed out in \cite{Silverstein&Follum:20NAPSI}, data compression is needed for real-time streaming and event-specific polling.

This paper presents a practical solution derived from a low complexity technology that has been widely used outside the power system domain.  The developed technique applies to the streaming of both CPOW and PMU measurements in either on-demand/event-driven or continuous streaming modes.   The main contribution of this work lies in specializing subband compression techniques to the power system-specific decomposition of signal bandwidth into harmonic subbands.  To this end, we demonstrate that ASBC can achieve 100 to 1000:1 compression ratios necessary to monitor a wide range of transient events.

{We have compared the Fourier-based  (ASBC and FFT) and wavelet-based (DWT) compression schemes using a combination of synthetic and real data, albeit in limited laboratory settings. In these experiments,  ASBC performed better than generic implementations of FFT and DWT-based compressions.  DWT-based techniques are known for their advantages of capturing waveform details, and  there is a significant literature  on the effectiveness of DWT-based compression when the compression ratio is relatively low.  For CPOW streaming with high compression ratio, however,  DWT-based techniques did not offer sufficiently high compression ratio for the level of  accuracy required by industrial standards in our experiments. Such characteristics need to be validated in more extensive field studies.}

We have left out several issues  that would be of significance in future work.   ASBC, in its current form, applies to compressions at individual CPOW sensors.  Power system measurements are constrained by physical laws that impose strong spatial dependencies.   A natural framework is to exploit spatial dependencies at data concentrators.   In such settings,  ASBC can be combined with some of the spatiotemporal compression techniques  \cite{Gadde&etal:16TSG,Kummerow&Nicolai&Bretschneider:18EnergyCon} for either efficient storage or streaming. An issue of particular significance is data security and anomaly detection in CPOW data.  It worths mentioning that CPOW data, unlike PMU and SCADA measurements, are more difficult for data injection attacks \cite{Kosut&Jia&Thomas&Tong:10TSG}  and potentially easier to detect malicious data attack.

{
\bibliographystyle{IEEEtran}
\bibliography{BIB_PMU}

% Generated by IEEEtran.bst, version: 1.14 (2015/08/26)
\begin{thebibliography}{10}
\providecommand{\url}[1]{#1}
\csname url@samestyle\endcsname
\providecommand{\newblock}{\relax}
\providecommand{\bibinfo}[2]{#2}
\providecommand{\BIBentrySTDinterwordspacing}{\spaceskip=0pt\relax}
\providecommand{\BIBentryALTinterwordstretchfactor}{4}
\providecommand{\BIBentryALTinterwordspacing}{\spaceskip=\fontdimen2\font plus
\BIBentryALTinterwordstretchfactor\fontdimen3\font minus
  \fontdimen4\font\relax}
\providecommand{\BIBforeignlanguage}[2]{{%
\expandafter\ifx\csname l@#1\endcsname\relax
\typeout{** WARNING: IEEEtran.bst: No hyphenation pattern has been}%
\typeout{** loaded for the language `#1'. Using the pattern for}%
\typeout{** the default language instead.}%
\else
\language=\csname l@#1\endcsname
\fi
#2}}
\providecommand{\BIBdecl}{\relax}
\BIBdecl

\bibitem{Liu:20Workshop}
Y.~Liu, ``Beyond today's synchrophasor,'' in \emph{NSF Workshop on Forging
  Connections between Machine Learning, Data Science, and Power Systems
  Research}, March 2020, [ONLINE], available (2020/8/20) at
  \url{https://sites.google.com/umn.edu/ml-ds4pes/presentations}.

\bibitem{Silverstein&Follum:20NAPSI}
A.~Silverstein and J.~Follum, ``High-resolution, time-synchronized grid
  monitoring devices,'' North American Synchrophasor Initiative, Tech. Rep.
  NAPSI-2020-TR-004, March 2020, [ONLINE], available (2020/8/20) at
  \url{https://www.naspi.org/node/819}.

\bibitem{Anderson&Agrawal&vanNess:99book}
P.~M. Anderson, B.~L. Agrawal, and J.~E.~V. Ness, \emph{Subsynchronous
  Resonance in Power Systems}.\hskip 1em plus 0.5em minus 0.4em\relax
  Wiley-IEEE Press, 1999.

\bibitem{Perez:10PRE}
J.~{Perez}, ``A guide to digital fault recording event analysis,'' in
  \emph{63rd Annual Conference for Protective Relay Engineers}, 2010, pp.
  1--17.

\bibitem{NERC:17}
{NERC}, ``1200 mw fault induced solar photovoltaic resource interruption
  disturbance report: Southern california 8/16/2016 event,'' North American
  Electric Reliability Corporation, Tech. Rep., June 2017, [ONLINE], available
  (2021/1/1) at
  \url{https://www.nerc.com/pa/rrm/ea/Pages/1200-MW-Fault-Induced-Solar-Photovoltaic-Resource-Interruption-Disturbance-Report.aspx}.

\bibitem{Tcheou&etal:14TSG}
M.~P. {Tcheou} \emph{et~al.}, ``The compression of electric signal waveforms
  for smart grids: State of the art and future trends,'' \emph{IEEE
  Transactions on Smart Grid}, vol.~5, no.~1, pp. 291--302, Jan 2014.

\bibitem{Cover&Thomas:Book}
T.~Cover and J.~Thomas, \emph{Elements of Information Theory}.\hskip 1em plus
  0.5em minus 0.4em\relax John Wiley \& Sons, Inc., 1991.

\bibitem{Mehta&Russell:89TPD}
K.~{Mehta} and B.~D. {Russell}, ``Data compression for digital data from power
  systems disturbances: requirements and technique evaluation,'' \emph{IEEE
  Transactions on Power Delivery}, vol.~4, pp. 1683--1688, 1989.

\bibitem{Littler&Morrow:99TPD}
T.~B. {Littler} and D.~J. {Morrow}, ``Wavelets for the analysis and compression
  of power system disturbances,'' \emph{IEEE Transactions on Power Delivery},
  vol.~14, no.~2, pp. 358--364, April 1999.

\bibitem{Santoso&Powers&Grady:97TPD}
S.~{Santoso}, E.~J. {Powers}, and W.~M. {Grady}, ``Power quality disturbance
  data compression using wavelet transform methods,'' \emph{IEEE Transactions
  on Power Delivery}, vol.~12, no.~3, pp. 1250--1257, July 1997.

\bibitem{Gaouda&etal:00TPD}
A.~M. {Gaouda}, M.~M.~A. {Salama}, M.~R. {Sultan}, and A.~Y. {Chikhani},
  ``Application of multiresolution signal decomposition for monitoring
  short-duration variations in distribution systems,'' \emph{IEEE Transactions
  on Power Delivery}, vol.~15, no.~2, pp. 478--485, April 2000.

\bibitem{Santoso&etal:00TPD}
S.~{Santoso}, E.~J. {Powers}, W.~M. {Grady}, and A.~C. {Parsons}, ``Power
  quality disturbance waveform recognition using wavelet-based neural
  classifier. i. theoretical foundation,'' \emph{IEEE Transactions on Power
  Delivery}, vol.~15, no.~1, pp. 222--228, Jan 2000.

\bibitem{Dash&etal:03TPD}
P.~K. {Dash}, B.~K. {Panigrahi}, D.~K. {Sahoo}, and G.~{Panda}, ``Power quality
  disturbance data compression, detection, and classification using integrated
  spline wavelet and s-transform,'' \emph{IEEE Transactions on Power Delivery},
  vol.~18, no.~2, pp. 595--600, April 2003.

\bibitem{Tse&etal:12TPD}
N.~{Tse} \emph{et~al.}, ``Real-time power-quality monitoring with hybrid
  sinusoidal and lifting wavelet compression algorithm,'' \emph{IEEE
  Transactions on Power Delivery}, vol.~27, no.~4, pp. 1718--1726, Oct 2012.

\bibitem{Meher&Pradhan&Panda:03EPSR}
S.~Meher, A.~Pradhan, and G.~Panda, ``An integrated data compression scheme for
  power quality events using spline wavelet and neural network,''
  \emph{Electric Power Systems Research}, vol.~69, 2004.

\bibitem{Ning&Etal:11TSG}
J.~{Ning}, J.~{Wang}, W.~{Gao}, and C.~{Liu}, ``A wavelet-based data
  compression technique for smart grid,'' \emph{IEEE Transactions on Smart
  Grid}, vol.~2, no.~1, pp. 212--218, March 2011.

\bibitem{Ibrahim&Morcos:05TPD}
W.~R.~A. {Ibrahim} and M.~M. {Morcos}, ``Novel data compression technique for
  power waveforms using adaptive fuzzy logic,'' \emph{IEEE Transactions on
  Power Delivery}, vol.~20, no.~3, pp. 2136--2143, 2005.

\bibitem{Ge&etal:15TSG}
Y.~{Ge}, A.~J. {Flueck}, D.~{Kim}, J.~{Ahn}, J.~{Lee}, and D.~{Kwon}, ``Power
  system real-time event detection and associated data archival reduction based
  on synchrophasors,'' \emph{IEEE Transactions on Smart Grid}, vol.~6, no.~4,
  pp. 2088--2097, July 2015.

\bibitem{Klump&etal:10PESGM}
R.~{Klump}, P.~{Agarwal}, J.~E. {Tate}, and H.~{Khurana}, ``Lossless
  compression of synchronized phasor measurements,'' in \emph{IEEE PES General
  Meeting}, 2010, pp. 1--7.

\bibitem{Gadde&etal:16TSG}
P.~H. {Gadde}, M.~{Biswal}, S.~{Brahma}, and H.~{Cao}, ``Efficient compression
  of pmu data in {WAMS},'' \emph{IEEE Transactions on Smart Grid}, vol.~7,
  no.~5, pp. 2406--2413, 2016.

\bibitem{Kummerow&Nicolai&Bretschneider:18EnergyCon}
A.~{Kummerow}, S.~{Nicolai}, and P.~{Bretschneider}, ``Spatial and temporal
  {PMU} data compression for efficient data archiving in modern control
  centres,'' in \emph{2018 IEEE International Energy Conference (ENERGYCON)},
  2018, pp. 1--6.

\bibitem{Das&Sidhu:14TII}
S.~{Das} and T.~{Singh Sidhu}, ``Application of compressive sampling in
  synchrophasor data communication in {WAMS},'' \emph{IEEE Transactions on
  Industrial Informatics}, vol.~10, no.~1, pp. 450--460, Feb 2014.

\bibitem{Zhang&etal:15TPS}
F.~{Zhang} \emph{et~al.}, ``Application of a real-time data compression and
  adapted protocol technique for {WAMS},'' \emph{IEEE Transactions on Power
  Systems}, vol.~30, no.~2, pp. 653--662, March 2015.

\bibitem{Testa&etal:07TPD}
A.~{Testa} \emph{et~al.}, ``Interharmonics: Theory and modeling,'' \emph{IEEE
  Transactions on Power Delivery}, vol.~22, no.~4, pp. 2335--2348, 2007.

\bibitem{IEEEStdC37:11}
IEEE, ``I{EEE} standard for synchrophasor data transfer for power systems,''
  \emph{IEEE Std C37.118.2-2011 (Revision of IEEE Std C37.118-2005)}, pp.
  1--53, Dec 2011.

\bibitem{Kosut&Jia&Thomas&Tong:10TSG}
O.~{Kosut}, L.~{Jia}, R.~J. {Thomas}, and L.~{Tong}, ``Malicious data attacks
  on the smart grid,'' vol.~2, no.~4, 2011, pp. 645--658.

\end{thebibliography}
}

%\newpage
%\section*{Appendix}
%\input Appendix_v7

\end{document}